\documentclass[12pt]{article}
\usepackage{indentfirst,psfig}
\newcommand{\bey}[1]{\begin{eqnarray} \label{#1}}
\newcommand{\eey}{\end{eqnarray}}
\newcommand{\beq}[1]{\begin{equation} \label{#1}}
\newcommand{\eeq}{\end{equation}}
\makeatletter
\def\@evenhead{\vbox{\hbox to\hsize{\sl \headmsg \hfill}}}
\def\@oddhead{\vbox{\hbox to\hsize{\sl \headmsg \hfill }}}
\makeatother
\def\headmsg{}

\title{A Hierarchical Approach to Protein Molecular Evolution}

\author{ Leonard D. Bogarad$^*$ and Michael W. Deem$^\dagger$\\
~\\
\hbox{}$^*$Division of Biology, California Institute of Technology\\
Pasadena, CA  91125\\
~\\
\hbox{}$^\dagger$Chemical Engineering Department, University of California\\
Los Angeles, CA 90095--1592}

\begin{document}
\maketitle
\renewcommand{\baselinestretch}{1.3}
\tiny
\normalsize

{\flushleft


\emph{Proc.\ Natl.\ Acad.\ Sci.\ USA}, to appear.
\bigskip

Physical Sciences. Chemistry.

Biological Sciences. Evolution.
}

\newpage
\def\headmsg{L. D. Bogarad and M. W. Deem, ``A Hierarchical Approach
to Protein Molecular Evolution''}

\centerline{\textbf{ABSTRACT}}
\begin{quote}
Biological diversity has evolved despite the essentially infinite
 complexity of protein sequence
space. We present a hierarchical approach to the efficient searching of
 this space and
quantify the evolutionary potential of our approach with Monte Carlo
 simulations. These
simulations demonstrate that non-homologous juxtaposition of encoded
 structure is the
rate-limiting step in the production of new tertiary protein
 folds.  Non-homologous ``swapping'' of
low energy secondary structures increased the binding constant
 of a simulated protein by $\approx10^7$
relative to base substitution alone. Applications of our approach include
the generation
of new protein folds and modeling the molecular evolution of disease.
\end{quote}

\newpage

The exponential complexity of protein space limits evolution via DNA
base substitution alone and remains a major challenge to many
quantitative treatments of evolution. 
Random assembly and base
substitution are ideally suited for searching local regions of
polypeptide space, as demonstrated experimentally by the isolation of
large numbers of stable structures from random encoded peptide
libraries 
\cite{len1,len2,len3} 
and the rapid improvement of function
seen in molecular evolutions of synthetic antibodies
\cite{len4,len5,len6}.
 However, \emph{in vitro} homologous recombination
experiments, termed DNA shuffling, have already demonstrated the
limitations of protein evolution via base substitution alone
\cite{len7,len8,len9,len10,len11}. Indeed, a complete hierarchy of natural
mutational events composed of rearrangements, deletions, horizontal
transfers \cite{len12}, transpositions \cite{len13}, and other
non-homologous juxtapositions, in addition to base substitution and
homologous recombination, is required for the rapid generation of
protein diversity.

Modern neo-Darwinism and neutral evolutionary treatments, therefore, fail to
explain satisfactorily the generation of the diversity of life found
on our planet. Yet most theoretical treatments of evolution
consider only the limited point mutation events that form the
basis of these theories.  Similarly, methods of experimental
protein evolution are generally
 limited to point mutation and DNA shuffling.  Genetic studies, on the
other hand, have indicated the importance
of dramatic, DNA swapping events in natural evolution
\cite{len12,len28,len29,len30,Shapiro,Shapiro2}.

We here address from a theoretical point of view the question of how 
protein space can be searched efficiently and thoroughly, either
in the laboratory or in Nature.  
We demonstrate that
point mutation alone is incapable of evolving systems with
substantially new protein folds. 
We further demonstrate that even the DNA shuffling approach is
incapable of evolving substantially new protein folds.
  Our
Monte Carlo simulations demonstrate that non-homologous DNA ``swapping''
of low energy structures is a key step in searching
protein space.  

More generally, our simulations demonstrate
that the efficient search of large regions of protein
space requires a hierarchy of genetic events, each encoding higher
order structural substitutions.  
We show how the complex protein
function landscape can be navigated with these moves.  
We conclude
that analogous moves have driven the evolution of protein
diversity found in Nature. 
We suggest that our moves, which appear to be experimentally feasible,
would make an interesting addition to the techniques of molecular
biotechnology.  Applications of
our approach include improvement of current molecular evolution
techniques, generation of non-natural protein folds, and modeling
the molecular evolution of disease.

{\bf The Generalized Block NK Model}.
We performed model Monte Carlo simulations to quantify and
optimize hierarchical protein space searching by genetic means.
Molecular evolution strategies were simulated using an energy
function as the selection criterion. The
energy function  is a generalization of the NK
\cite{len14,len15,len16} and block NK \cite{len17} models. Our energy
function takes into account the spontaneous generation of convergent
secondary structures via the interactions of amino acid side chains
 as well as the interactions between secondary structures
within proteins. In addition, we include a contribution
to model binding to a substrate. This approach assigns a
unique energy value to each evolving protein sequence. This model,
while a simplified description of real proteins, captures much of the
thermodynamics of protein folding and ligand binding.
This generalized NK model contains several parameters, and a reasonable
determination of these parameters is what allows the model to compare successfully 
with experiment.
 The combined
ability to fold and bind substrate is what we seek to optimize. 
That is, the direction of our protein evolution will be based upon
this energy function.

The specific energy function used as the selection criterion in our
molecular simulations is
\beq{101}
 U = \sum_{\alpha = 1}^M U_\alpha^\mathrm{sd} +
\sum_{\alpha > \gamma = 1}^M U_{\alpha \gamma}^\mathrm{sd-sd}
+ \sum_{i = 1}^P U_i^\mathrm{c} \ .
\eeq
This energy function is composed of three parts: secondary structural
subdomain energies ($U^\mathrm{sd}$), subdomain-subdomain interaction energies
($U^\mathrm{sd-sd}$), and chemical binding energies ($U^\mathrm{c}$).  Each of these
three energy terms is weighted equally, and each has a magnitude near
unity for a random sequence of amino acids. In this NK based
simulation, each different type of amino acid behaves as a completely
different chemical entity; therefore, only five chemically distinct
amino classes are considered (\emph{e.g.}, negative, positive, polar,
hydrophobic, and other). Simplified amino acid alphabets not only
are capable of producing functional proteins 
\cite{len19,len20} 
but also
may have been used in the primitive genetic code
 \cite{len21,len22}.
Simulated proteins have $M=10$ secondary structural subdomains of $N=10$
amino acids in length. They belong to one of $L=5$ different types
(\emph{e.g.}, helices, strands, loops, turns, and others). This gives
L different ($U^\mathrm{sd}$) energy functions of the NK form
\cite{len14,len15,len16,len17}. 
\beq{102}
U_\alpha^\mathrm{sd} = \frac{1}{\left[ M(N-K)\right]^{1/2}}
\sum_{j = 1}^{N-K+1} \sigma_\alpha \left(
a_j, a_{j+1}, \ldots, a_{j+K-1}
\right) \ .
\eeq
We consider $Q=5$ different chemical classes of amino acids with $K=4$
interactions \cite{len16}. The quenched, unit-normal random        
number $\sigma_\alpha$ in Eq.\ \ref{102} is different for each
value of its argument for each of the $L$ classes. This random form
mimics the complicated amino acid side chain interactions
within a given secondary structure. The energy of interaction
between secondary structures is given by
\bey{103}
U_{\alpha \gamma}^\mathrm{sd-sd} &=& \left[
\frac{2}{D M (M-1)} \right]^{1/2} \nonumber \\
&&\times 
\sum_{i=1}^D \sigma_{\alpha \gamma}^{(i)}
\left(
a_{j_1}^\alpha, \dots,
a_{j_{K/2}}^\alpha;
a_{j_{K/2+1}}^\gamma, \ldots
a_{j_{K}}^\gamma
 \right) \ .
\eey
We set the number of interactions between secondary structures at
$D=6$. Here the unit-normal weight, $\sigma_{\alpha \gamma}^{(i)}$,
and the interacting amino acids, $\{j_1,\ldots,j_K\}$, are selected at
random for each interaction $(i, \alpha, \gamma)$. The chemical
binding energy of each amino acid is given by
\beq{104}
U_i^\mathrm{c} = \frac{1}{\sqrt P} \sigma_i \left( a_i \right) \ .
\eeq
The contributing amino acid, $i$, and the unit-normal weight of the
binding, $\sigma_i$, are chosen at random.  
We assume $P=5$ amino acids contribute directly to the binding
event, as in a typical pharmacophore.

{\bf Simulation Protocol}.
In each simulated mutagenesis we started with 10,000 copies of a 100 amino acid
polypeptide sequence. We determined that it was optimal to keep the 10\%
best (lowest energy) protein sequences after selection and then amplify
these back up to a total of 10,000 copies
before the process was repeated. In each experiment 100 rounds of mutagenesis
and selection were
performed due to the relatively low optimal rates of base substitution
and DNA swapping.
To obtain a base line for searching fold space, we simulated molecular
evolution via simple mutagenesis (see Figure 1a).  Simulated
evolutions by amino acid substitution lead to significantly improved
protein energies. These evolutions always terminated at local energy
minima, however (see Table \ref{table1}).  This trapping is due to
the difficulty of combining the large numbers of
individual highly correlated substitutions necessary to generate new
protein folds.  Increasing the screening stringency in later rounds
did not improve the binding constants of simulated proteins, most
likely due to the lack of additional selection criteria such as
growth rates. Although we directly simulated only non-conservative
mutations, 
conservative and synonymous neutral mutations are not excluded and
could be taken into account in a more detailed treatment.
Indeed, our
optimized average mutation rate of 1 amino acid
substitution/sequence/round is equivalent to roughly 1-6 random base
substitutions/round.

{\bf Simulated DNA Shuffling}.
DNA shuffling improves the search of local fold space via a random yet
correlated combination of homologous coding fragments that contain limited
numbers of beneficial amino acid substitutions.  As in experimental
evolutions \cite{len7,len8,len9,len10}, the simulated shuffling
improved protein function significantly better than did point
mutation alone (see Table \ref{table1} and Figure 1b). However, local
barriers in the energy function 
also limit molecular evolution via DNA shuffling.
For example,  when we increased our screen size to 
20,000 proteins per round, we saw no
further improvement  in the final evolved energies.
Interestingly, our
optimal simulated DNA shuffling length of 20 amino acids (60 bases) is
nearly identical to fragment lengths used in experimental protocols
\cite{len8}.

{\bf Single-Pool Swapping}.
In Nature, local protein space can be rapidly searched by the directed
recombination of encoded domains from multi-gene pools.
A prominent example is the creation of the primary antibody
repertoir in an
adaptive immune system. We generalized these events by
simulating the swapping of amino acid fragments from 5 different
structural pools representing helices, strands, loops, turns, and
others (see Figure 1c).  During the swapping step, subdomains were
randomly replaced with members of the same secondary structural pools
with an optimal probability of 0.01/subdomain/round. We limited the
simulated evolution of the primary fold by maintaining the linear
order of swapped secondary structure types. The addition of the
swapping move was so powerful that we  were able to achieve binding
constants 2 orders of magnitude higher than in shuffling simulations
(see Table \ref{table1}).  Significantly, these improved binding
constants were achieved starting with
10-20 times less minimized structural subdomain material.

{\bf Parallel Experiments}.
Parallel tempering is a powerful statistical method that often allows
a system to escape local energy minima \cite{len23}.
This method simultaneously simulates several 
systems at different temperatures, allowing systems
at adjacent temperatures to swap configurations.
The swapping between high- and low-temperature systems allows for
an effective searching of configuration space.
  In Nature, as
well, it is known that genes, gene fragments, and gene operons are
transfered between species of different evolutionary complexity (\emph{i.e.},
at different ``temperatures'').   By analogy, we simulated
limited population mixing among parallel swapping experiments
by randomly exchanging evolving proteins at an optimal
probability of 0.001/protein/round. These mixing simulations optimized local
space searching and
achieved binding constants $\approx10^5$
higher than did base substitution alone (see Table \ref{table1}). Improved
function is due, in part, to the increased number of events in parallel
experiments.  Indeed, mixing may occur in Nature when
the evolutionary target function changes with time.
That is, in a dynamic environment
with multiple selective pressures, mixing would 
be especially effective when the rate of evolution of an isolated
population is slower than the rate of environmental change. 
Recently, it has been demonstrated that the mixing and DNA
shuffling of orthologous proteins resulted in rapid and dramatic increases
in recombinant protein function \cite{len8}.
It has also been argued recently that spatial heterogeneity in drug
concentration (a form of ``spatial parallel tempering'') facilitates the
evolution of drug resistance \cite{Perelson}.

{\bf Multi-Pool Swapping}.
The effective navigation of protein space requires the discovery and
selection of tertiary structures. To model the large scale search of
this space, we began with random polypeptide
sequences and repeated our swapping protocol, but now allowed secondary
structures from all 5 different pools to swap in at every position
(see Figure 1d).  This multi-pool swapping approach 
evolved proteins with binding constants $\approx10^7$ better than did
amino acid substitution of a protein-like starting sequence (see Table
\ref{table1}).  This evolution was accomplished by the
random yet correlated
juxtaposition of different types of low energy secondary structures.
 This approach dramatically improved specific
ligand binding while efficiently discovering new tertiary structures
(see Figure 2).  Optimization of the rate of these hierarchical molecular
evolutionary moves, including relaxation of
the selection criteria, enabled the
protein to evolve despite the high rate of failure for these dramatic
swapping moves.
 Interestingly, of all the molecular evolutionary
processes that we modeled, only multi-pool swapping demonstrated chaotic
behavior in
repetitive simulations.  This chaotic behavior was likely due to
the discovery of different model folds that varied
in their inherent ability to serve as scaffolds for ligand specific binding.

{\bf Possible Experimental Implementations}.
The search of large regions of protein space should identify new folds
and functions that would be of great value to basic, industrial, and
medical research. Our multi-pool searching protocol could be
attempted experimentally within present constraints 
($\approx10^4$--$10^{15}$ recombinants, depending on the screening or
selection method). One possibility is the combination of DNA
shuffling with synthetic splicing libraries 
\cite{len24} 
that contain
representative pools of native low energy structures encoded within
multiple ($\approx10$) short exons. Alternatively, it should be
possible to generate multiple libraries of synthetic oligonucleotide
pools 
\cite{len25,len26} 
encoding numerous specific secondary and
subdomain structures. Asymmetric, complementary encoded linkers with
embedded restriction sites would make the assembly, shuffling, and
swapping steps possible.

{\bf Parallels with Natural Evolution}.
During the course of any evolutionary process, proteins become trapped
in local energy
minima. Dramatic moves, such as swaps and juxtapositions, are needed to
break out of these
regions. Dramatic moves are usually deleterious, however. The evolutionary
success of these
events depends on population size; generation time; mutation rate;
population mixing;
selective pressure or freedom, such as successful genome
duplications or the
establishment of set-aside cells \cite{len27}; and the mechanisms that transfer
low energy, encoded structural domains.

In Nature, mechanisms have evolved to increase the probabilities of
successful exchanges. Viruses and transposons, for example, have
evolved large-scale integration mechanisms, while terminal VDJ
recombination is effective despite $> 50$\%\  in-frame failures.
While random swapping of genomic DNA is unlikely to lead to useful
protein products at a high rate,  a possible senario is that
 exon shuffling generated the primordial fold diversity
\cite{len28,len29,len30}.  This 
 hypothesis is bolstered by the correlation between
splice junction location and boundaries of encoded structural
domains.  Alternatively, if splicing was not
primitive, random swapping by horizontal transfer, rearrangement,
recombination, deletion, and insertion could have led to high
in-frame success rates if primitive genomes had high densities of
coding domains and reading frames, as in certain prokaryotes and
mitochondria. 

Three dramatic examples of use of swapping by Nature are particularly
notable.   The first is the development of antibiotic resistance.  It
was originally thought that no bacteria would become resistant to
penicillin due to the many point mutations required for resistance. 
Resistance occurred, however, within several years.  It is now known that
this resistance occurred through the swapping of pieces of DNA between
evolving bacteria \cite{Shapiro,Shapiro2}.  Multi-drug resistance is now a
major, current health care problem.  The creation of the 
primary antibody repertoire in vertebrates
 is another example of DNA swapping (of
genes, gene segments, or pseudo-genes).
  Finally, the evolution of \emph{E. coli} from
\emph{Salmonella} occurred exclusively by DNA swapping
\cite{len12}.  Indeed, none of the phenotypic differences between these two
species is due to point mutation. Moreover, even the observed
rate of evolution due to DNA swapping, 31,000 bases/million years,
is higher than that due to point mutation, 22,000 bases/million years.
That is, even though a DNA swapping event is less likely to be tolerated
than is a point mutation, the more dramatic nature of the swapping event
leads to a higher overall rate of evolution.  This is exactly the
behavior we observed in our simulations.

{\bf Summary}.
DNA base substitution, in the context of the genetic code, is
ideally suited for the
generation, diversification, and optimization of local protein space 
\cite{len21,len31}. However, the
difficulty of making the transition from one
productive tertiary fold to another limits
evolution via base
substitution and homologous recombination alone (Figure 2, light and
 dark yellow arrows,
respectively). Non-homologous DNA recombination, rearrangement, and
 insertion allow for
the combinatorial creation of productive tertiary folds via the novel
juxtaposition of suitable,
encoded structures. The efficient search of high-dimensional
fold space is dependent
upon the spontaneous generation and convergence of secondary structure
and the hierarchical
range of DNA mutation events present in our model (Figures 1d and 2,
purple arrow).
Starting with very small pools of low energy secondary structures, we
evolved new protein
folds with specific binding constants $\approx10^7$ higher than those optimized
by base substitution
alone. More generally, it seems likely that
organization into higher order fundamental units such as nucleic
acids, the genetic
code, secondary and tertiary structure, cellular compartmentalization,
cell types, and germ
layers allows systems to escape complexity barriers and
 potentiates explosions in diversity.

 Qualitative changes in
protein space such as those modeled here allow viruses, parasites,
bacteria, and cancers to evade the immune system, vaccines,
antibiotics, and therapeutics. The successful design of vaccines and
drugs must anticipate the evolutionary potential of both local and
large space searching by pathogens in response to therapeutic and
immune selection.  The addition of disease specific constraints to
our Monte Carlo simulations should be a promising approach for
predicting pathogen plasticity. Experimental implementation of our
hierarchical protocol should be a powerful approach to the discovery
of new therapeutics.  Infectious agents will continue to evolve
unless we can force them down the road to extinction.

{\bf Acknowledgments}.
We thank Daan Frenkel and Jonathan Rast for critical
readings of our manuscript.

\newpage

\bibliography{evolve}
\newpage

\begin{table}[htbp]
\caption{Results of Monte Carlo simulation of the evolution protocols.$^*$}
\label{table1}
\begin{center}
\begin{tabular}[t]{cccc}
\hline
\hline
Evolution Method &  Starting Energy&   Evolved Energy&    Achieved Binding Constant\\
\hline
Amino Acid Substitution &     -17.00 &        -23.18 &             1\\
DNA Shuffling        &   -17.00      &   -23.83      &        100\\
Swapping            &0       & -24.52             & $1.47 \times 10^4$\\
Mixing              &0       & -24.88           &   $1.81 \times 10^5$\\
Multi-Pool Swapping &      0 &       -25.40$^{~\dagger}$ &    $8.80 \times 10^6$$^{~\dagger}$\\
\hline
\hline
\end{tabular}
\end{center}
\hbox{}$^*$The starting polypeptide energy of -17.00 comes from
a protein-like sequence (minimized $U^\mathrm{sd}$), and 0 comes from a random
initial sequence of amino acids. The evolved energies and binding constants
are median values. The binding constants are calculated as $a e^{-b U}$,
where $a$ and $b$ are constants determined by normalizing the binding
constants achieved by point mutation and shuffling to 1 and 100, respectively.\\
\hbox{}$^\dagger$Note that the energies
and binding constants achieved via multi-pool swapping represents
typical best evolved protein folds.
\end{table}
\clearpage

\newpage
 \begin{center}
{\bf Figure Captions}
\end{center}       

Figure 1: Schematic diagram of the simulated molecular evolution
protocols. a) Simulation of molecular evolution via base substitution
(substitutions are represented by orange dots). b) Simulated DNA
shuffling showing the optimal fragmentation length of 2 subdomains. c)
The hierarchical optimization of local space searching: The 250
different sequences in each of the 5 pools (\emph{e.g.}, helices, strands,
turns, loops, and others) are schematically represented by different
shades of the same color. d) The multi-pool swapping model for
searching vast regions of tertiary fold space is essentially the same
as in the Figure 1c except that now
sequences from all 5 different structural pools can be swapped into
any subdomain.
Multi-pool swapping
allows for the formation of new tertiary structures by changing the
type of secondary structure at any position along the protein.
\bigskip

Figure 2: Schematic diagram representing a portion of the
high-dimensional protein composition space.  The three-dimensional energy
landscape of Protein Fold 1 (green) is shown in cutaway. The arcs
with arrowheads represent the ability of a given molecular evolution
process to change the composition and so to
traverse the increasingly large barriers in the energy function. The smallest
arc (light yellow) represents the ability to evolve improved fold
function via point mutation. Then in increasing order: DNA shuffling
(dark yellow), swapping (orange), and mixing (red). Finally, our
multi-pool swapping model allows an evolving system to move (purple
arc) to a different energy landscape representing a new tertiary fold
(bottom). With this model, functional tertiary fold space has a large,
yet manageable, number of dimensions. That is, in 100 amino acids we
assume ~10 secondary structures of ~5 types (we balance rare forms
with the predominance of strands and helices) roughly yielding the
potential for $\approx10^7$ basic tertiary folds in Nature. Clearly,
organization into secondary structural classes represents a dramatic
reduction in the realized complexity of sequence space (\emph{e.g.},
\emph{versus} 300 bases of open reading frame DNA, $\approx10^{170}$, or
 100 amino
acids with a 20 letter or 5 letter genetic code, $\approx10^{130}$ or
$\approx10^{70}$, respectively). 
\newpage
\psfig{file=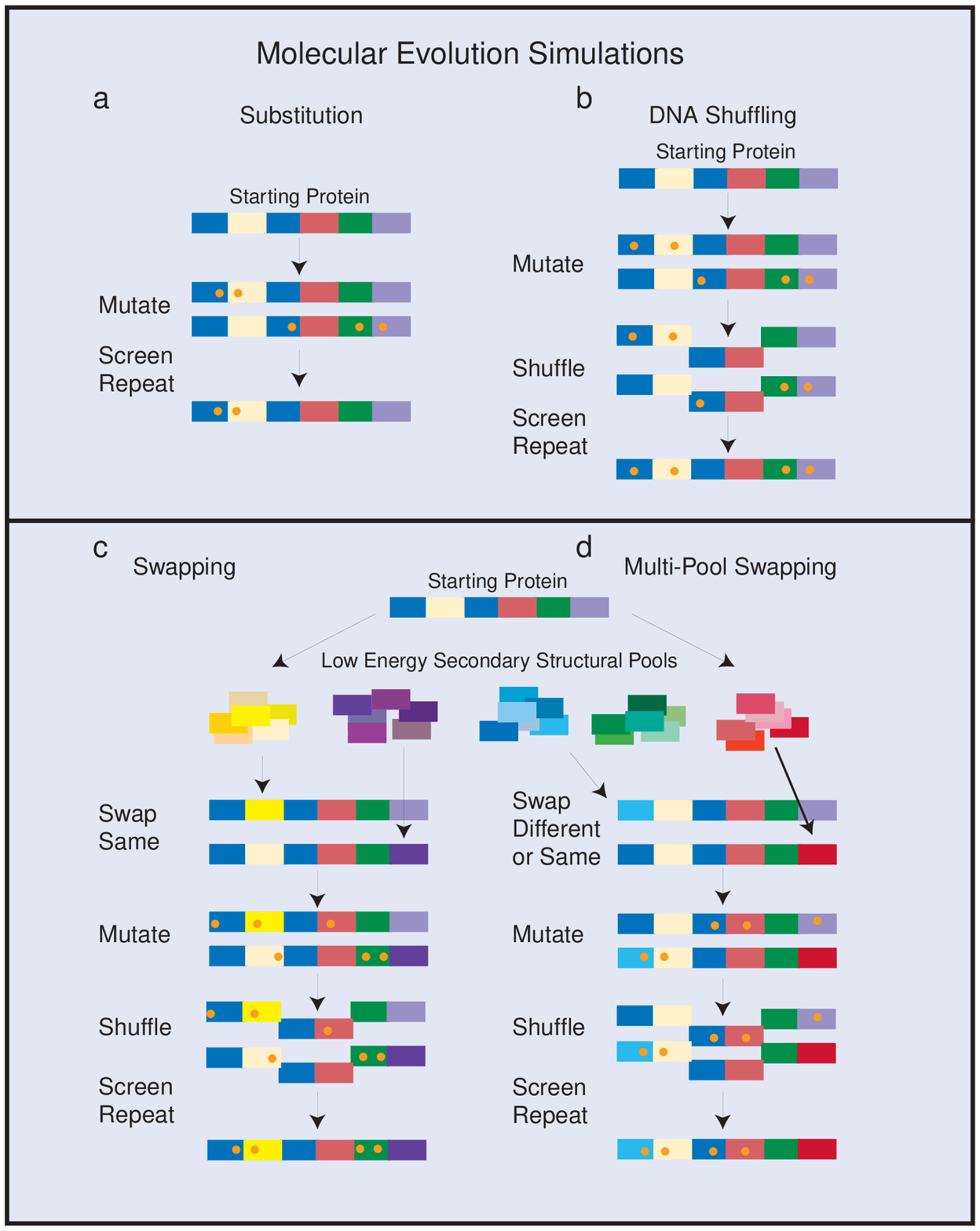,height=9in}
\clearpage
\psfig{file=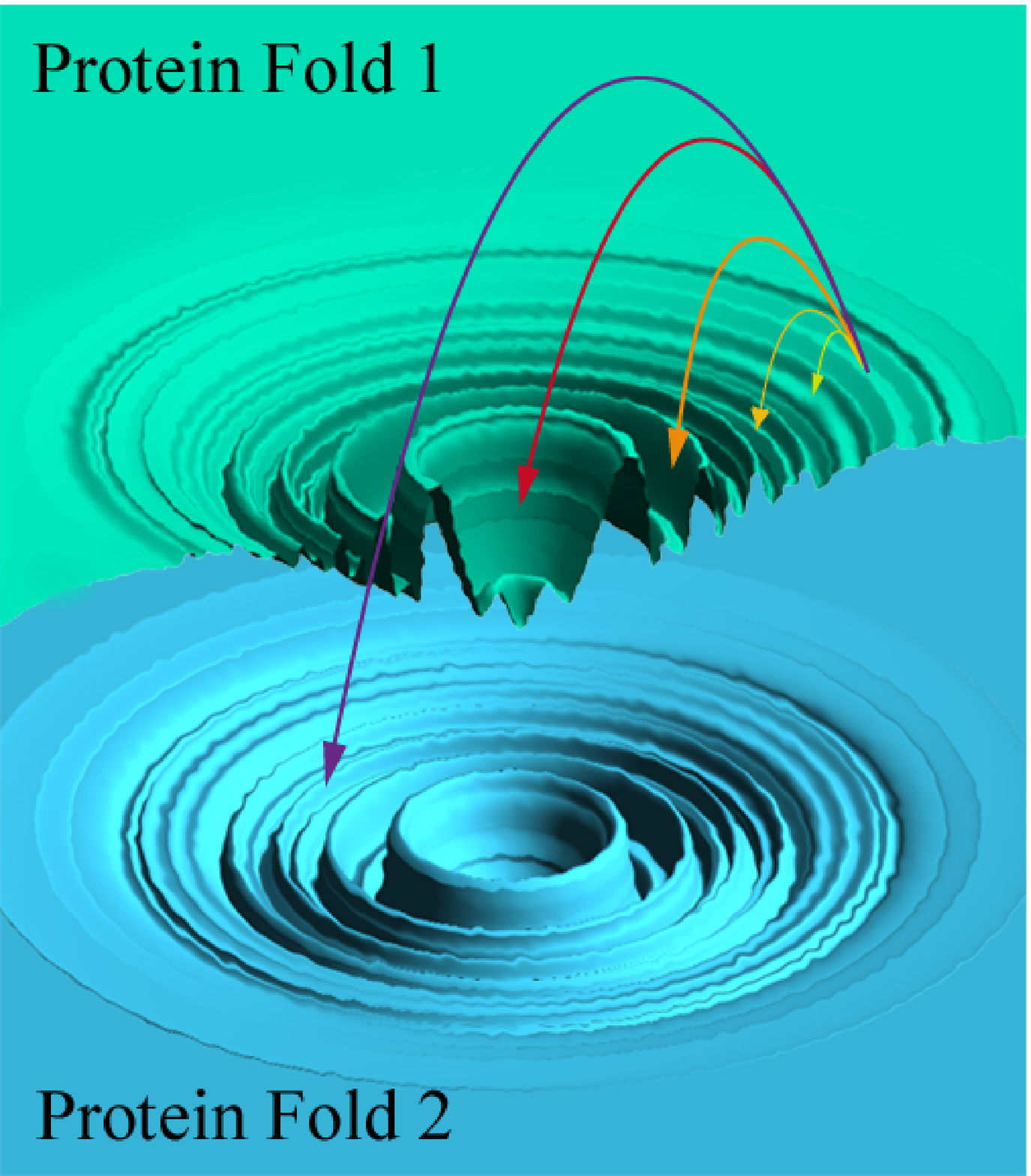,height=8in}
\end{document}